\documentclass[prd,twocolumn,superscriptaddress,showpacs,nofootinbib, amsmath, amssymb]{revtex4-1}

\usepackage{graphicx}
\usepackage{hyperref}
\usepackage{color}
\usepackage[utf8]{inputenc}

\usepackage{amssymb}
\usepackage{mathtools}
\newcommand{\convint}{\mathop{\mathrlap{\,\bigstar}\int}}

\newcommand{\m}[1]{{\bf \boldmath #1}} 

\newcommand{\units}{\ensuremath{\mathrm{cm}^{-2}\;\mathrm{s}^{-1}\;\mathrm{sr}^{-1}\;\mathrm{GeV}^{-1}}}

\begin{document}

\title{One-point fluctuation analysis of IceCube neutrino events outlines a significant unassociated isotropic component and constrains the Galactic contribution}

\author{Michael R. Feyereisen}
\affiliation{GRAPPA Institute, Institute of Physics, University of Amsterdam, 1098 XH Amsterdam, The Netherlands}
\author{Daniele Gaggero}
\affiliation{GRAPPA Institute, Institute of Physics, University of Amsterdam, 1098 XH Amsterdam, The Netherlands}
\author{Shin'ichiro Ando}
\affiliation{GRAPPA Institute, Institute of Physics, University of Amsterdam, 1098 XH Amsterdam, The Netherlands}
\affiliation{Kavli Institute for the Physics and Mathematics of the Universe (Kavli IPMU, WPI), Todai Institutes for Advanced Study, University of Tokyo, Kashiwa 277-8583, Japan}

\begin{abstract}
The origins of the extraterrestrial neutrinos observed in IceCube have yet to be determined. In this study we perform a one-point fluctuation analysis of the six-year high-energy starting event shower data,
with fixed non-Poissonian contributions from atmospheric, Galactic and some extragalactic components, as well as an isotropic (and weakly non-Poissonian) template. 
In addition to the star-forming galaxies and blazars, our analysis suggests the presence of an additional isotropic component, not associated with any detected class of point sources, with best-fit intensity of $(2.8\pm0.2)\times 10^{-18}\,(E/100~{\rm TeV})^{-2.7\pm 0.5}\;\units$.
For the first time, we also consider high-energy extrapolations of several phenomenological models for the diffuse Galactic emission (tuned to both local cosmic-ray data and diffuse gamma-ray emission in the GeV--TeV domain). We demonstrate the potential of our framework in discriminating between different scenarios, with possible implications on the physics of cosmic ray transport in the TeV--PeV range.
\end{abstract}

\maketitle

\section{Introduction}

A major recent breakthrough in the field of astroparticle physics is the detection of cosmic neutrinos by the IceCube collaboration. Four years after this epochal discovery, we are still far from understanding their origin. Given the spatial distribution, still consistent with isotropy~\cite{Aartsen:2016xlq, IceCubeICRC}, it is natural to assume that most events are extra-Galactic. Moreover, no significant clustering has been identified yet~\cite{IceCubePS}.

The community is now heavily debating which classes of sources contribute the most to the total flux, and what is the role of the Galactic component {(see, e.g. the constraints published by the ANTARES and IceCube collaborations \cite{ANTARESgal, IceCubegal}, and the analyses published in \cite{Ahlers2016prd,Pagliaroli2016jcap,Pagliaroli2017arxiv,Neronov2016app,Denton2017jcap,Palladino2016jcap,Palladino2016apj})}. In both issues the gamma-ray data clearly play a crucial role, and it is compelling to effectively exploit all the information we can obtain from gamma rays, and the diffuse Galactic emission data.

This paper addresses both questions at the same time by means of a one-point fluctuation analysis that extends Ref.~\cite{Feyereisen2016}. The technique maximally utilizes the statistical information contained in single pixels of the IceCube all-sky data. The same approach has proven to be powerful in constraining gamma-ray sources, both theoretically~\cite{Lee:2008fm, Feyereisen2015} and observationally~\cite{Zechlin:2016pme, Lisanti:2016jub} (see also Refs.~\cite{Ando:2017alx,Ando:2017xcb} for a complementary approach using the angular power spectrum).

\begin{figure*}
\begin{center}
\includegraphics[height=8cm]{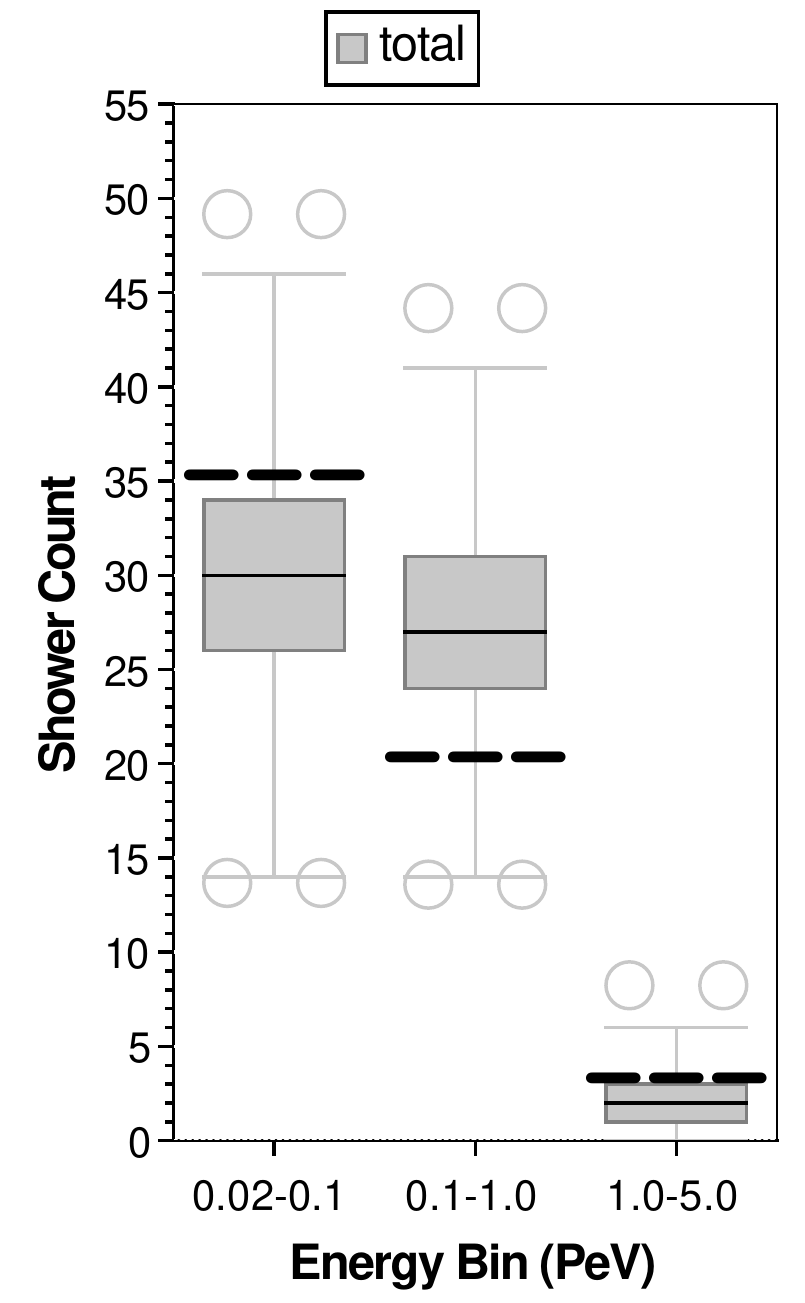}\hspace{5mm}
\includegraphics[height=8cm]{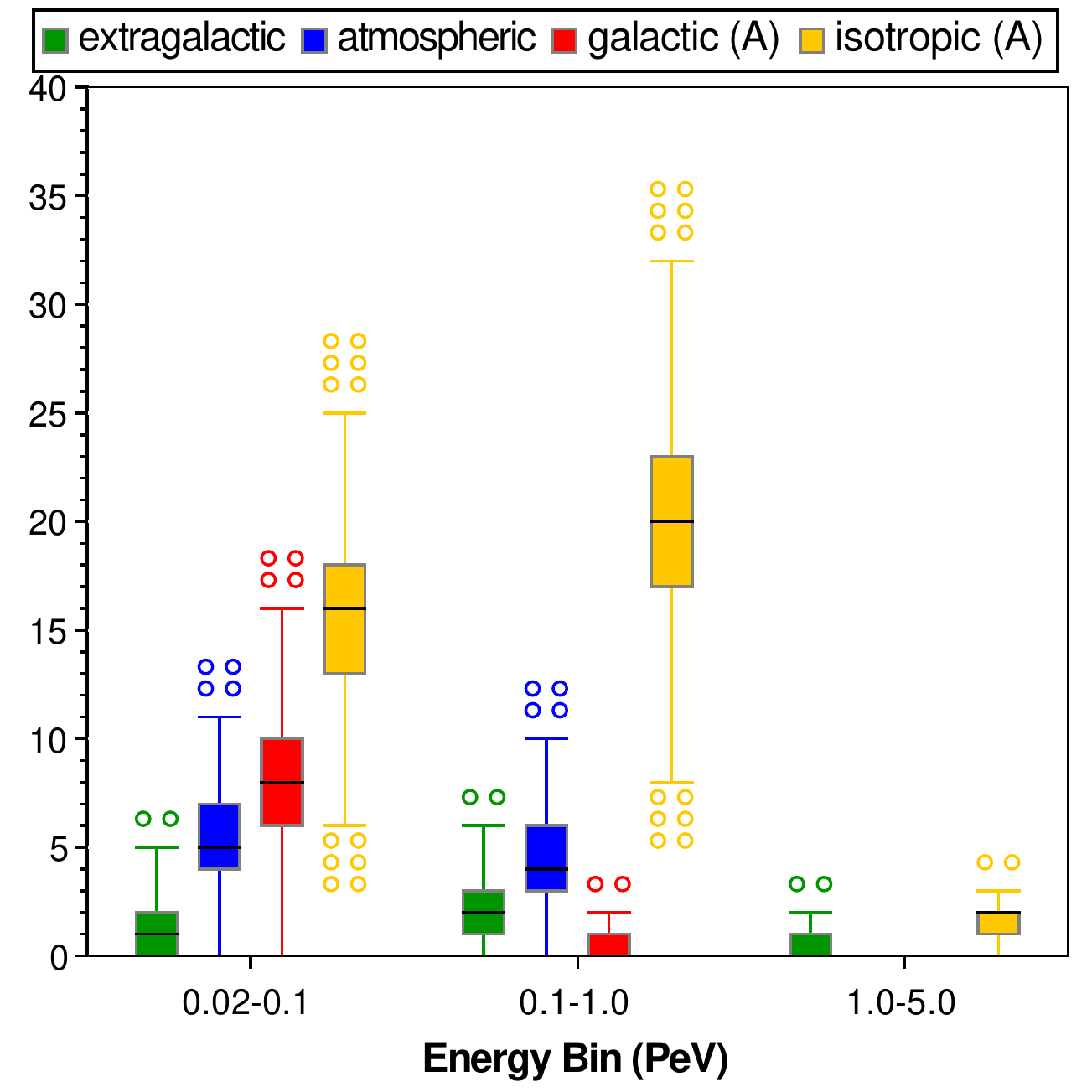}
\caption{{\itshape Left}: Whiskered-box plot of the aggregate and full-sky predicted HESE shower count distribution in each energy bin of the analysis. 50\% of predicted outcomes in each energy bin are contained in the solid box around the median, while whiskers show 1.5 times {the range covered by the box}. The number of small circles above and below the whiskers is in proportion to the number of outliers from simulations. The dashed line in each bin represents the actual observation. {\itshape Right}: The same as the left panel but showing individual predicted atmospheric and astrophysical contributions to the HESE shower events. 
In both panels, the assumed Galactic model and associated best-fit isotropic component are {\itshape Canonical} with low-energy cutoff (model \m{A}).}
\label{fig:all_energies_boxplot}
\end{center}
\end{figure*}

We apply this method to the latest IceCube shower data, specifically to six years of High-Energy Starting Events (HESE)~\cite{IceCubeICRC2017}.
We consider a comprehensive set of source classes.  Besides the fixed starburst and blazar templates considered in Ref.~\cite{Feyereisen2016}, we include -- for the first time in this context -- several models of the Galactic component originated from cosmic ray hadronic interactions, and a (hypothetical) additional, purely isotropic component.  For each of these components we compute the probability distribution function (PDF) of neutrino counts per IceCube pixel, which form the basis of a maximum likelihood analysis. 

Our main result is shown in Fig. \ref{fig:all_energies_boxplot} and clearly shows that the neutrino sky appears dominated by the isotropic component of unknown origin, and is compatible with a non-negligible Galactic component.

\section{Analysis}

\subsection{Method}

{The methodology applied in this work is the {\it one-point fluctuation analysis} (pioneered in \cite{1957PCPS...53..764S,1992ApJ...396..460B}, and recently applied to high-energy astrophysical data e.g. in \cite{Lee:2008fm,Feyereisen2015,Feyereisen2016}). 

This method allows to predict the one-point (pixel-by-pixel) neutrino count probability distribution for the set of source classes under investigation. 
This requires to:
\begin{itemize} 
\item Compute the intensity distributions $P(I_\nu)$, given a phenomenological or data-driven model for the classes of sources that are expected to contribute to the neutrino sky. In our case the model is based on multi-messenger data, and is mainly characterized by a luminosity function and a spectral template;
\item Convolve the distributions with the IceCube exposure (as a function of energy, flavor, and declination) for an observation period of six years. 
\end{itemize}

This procedure has already been applied in this context in Ref.~\cite{Feyereisen2016} Here we follow the same prescriptions for the pixelization and energy binning. However, we significantly expand the analysis by implementing an isotropic unassociated component (fitted to the neutrino data) and several models for the Galactic contribution tuned on gamma-ray and local cosmic-ray data. We note that our model is based on a {\it fixed} prediction for the neutrino count distribution of the classes of sources we consider, based on gamma-ray and infrared data; the {\it unassociated} component is the only one allowed to vary and fitted to the neutrino data themselves.}

In the next subsections we explain in the detail the data sample we use and the different astrophysical ingredients. {We refer the reader to the Appendix for more technical details about the procedure.}

\subsection{The HESE data sample}

The IceCube collaboration classifies neutrino events as having either a shower-like or track-like topology, the latter being a smoking gun of muonic interactions. The atmospheric background consists of not only atmospheric muon-neutrinos ($\nu_\mu$) but also atmospheric muons, some of which pass the stringent background-removal veto due to their sheer abundance.

In this study we focus on HESE, in particular those with the shower topology, since for this subsample the atmospheric $\nu_\mu$ contamination is minimized, and also because we do not need to worry about veto-passing muons. Our sample consists of the 58 shower events included in the six-year HESE data~\cite{IceCubeICRC2017}, three of which have energies above $1\;\mathrm{PeV}$. {This sample is the only one used in the analysis; however, we remark that in Fig. \ref{fig:galactic_boxplot} a subset of this data sample is visualized, with additional angular and energy cuts applied, as detailed in the caption.}

\subsection{The atmospheric foregrounds}

{The atmospheric neutrino flux has been measured very precisely for lower energies, and then extrapolated to the energy region of interest for this work.
Since physical processes of producing the atmospheric neutrinos are relatively well understood, we do not include any uncertainties related to the extrapolation.
We employ the average conventional atmospheric flux given by Ref.~\cite{Honda:2015fha} as $1.77\times 10^{-14}\;\units$ with a flavour ratio of $1:35:0$. Other percent-level atmospheric contributions from $\nu_e$ and $\nu_\tau$ fluxes~\cite{Honda:2015fha,Enberg:2008te} are neglected, as are the neutrino-antineutrino ratios, although the fully detailed (even energy-dependent) flavour ratios can in principle be accounted for in this type of analysis.
The prompt atmospheric neutrino flux is taken from Ref.~\cite{Enberg:2008te}, with a flavour ratio of $1:1:0$.

We do not take veto-passing muons into account in our analysis.
Since we focus only on shower events and the veto-passing muons are problematic only for track events, this is justified.
Accounting for tracks, on the other hand, would require adding time binning (to capture the seasonal variation of the atmospheric temperature) and a more involved modeling of the width of the atmospheric PDF to the analysis, both of which are beyond the scope of the present work.
}

\subsection{Extragalactic components}

{We now introduce our model for the neutrino emission from star-forming galaxies and blazars, which follows the prescriptions described in ~\cite{Feyereisen2016}.

\subsubsection{Star-forming galaxies}

Among star forming galaxies, we consider starburst galaxies (SBs) -- i.e. galaxies undergoing a short-duration exceptionally high rate of star formation -- and star-forming galaxies hosting an obscured or low-luminosity AGN (SF-AGN), as main contributors for the neutrino flux.

Since neutrino oscillations push the flavour ratio towards $1:1:1$, for all sources based on hadronuclear interactions we can define a general conversion between the all-flavour neutrino and antineutrino differential flux and the gamma-ray flux that simply reads $F_{\nu} = 6 F_{\gamma}$.
However, since star-forming galaxies are barely resolved in gamma rays, we do not rely on gamma-ray data, and choose to take into account the infrared luminosity function from the Herschel catalog instead. 
We then consider the empirical relation~\cite{2012ApJ...755..164A}
\begin{equation}
L_{\gamma} (L_{\rm IR}) \,=\, 10^\beta \, \left( \frac{L_{\rm IR}}{10^{10} \, L_{\odot}} \right)^{\alpha} \, \frac{\rm erg}{\rm s},
\end{equation}
where $\alpha \simeq 1.17$ and $\beta \simeq 39.3$, and convert the infrared luminosity function $\Phi_{\rm IR}(L_{\rm IR}, z)$ into a gamma-ray (and subsequently neutrino) luminosity function $\Phi_\gamma$ ($\Phi_\nu$).

Concerning the spectrum, we assume a fixed slope $\Gamma_{SB} = 2.2$. We also investigate the case of $\Gamma_{SB}=2.3$, which might be slightly more favoured~\cite{Bechtol:2015uqb}, but find that our conclusions do not change.

Once these ingredients are fixed, we can compute the gamma-ray flux distribution under the assumption that the sources are isotropically distributed in a comoving cosmological volume element~\cite{Feyereisen2016}:
\begin{equation}
P(F_{\gamma} \vert E_{\gamma}, \Gamma) \,\propto \, \frac{1}{F_{\gamma}} \int{ {\rm d}z \frac{{\rm d}V}{{\rm d}z} \frac{\Phi_\gamma(L_{\rm crit}, z)}{N \, {\rm ln}(10)} }.
\end{equation}

\subsubsection{Blazars} 

Blazars (or BL Lac objects) are jetted active galactic nuclei, with the jet pointed towards the observer. 
We first develop a gamma-ray model for this class of sources, relying on the the source count distribution inferred from the Second Catalog of Hard Fermi-LAT Sources (2FHL)~\cite{Ackermann:2015uya}, and assuming a spectral slope $\Gamma_{\rm 2FHL} = 2.5$ (which results in an optimistic estimate of their contribution, given the evidence that the blazar index is $\Gamma > 3$ at higher energies). 

Once the gamma-ray model is specified, we  exploit the following relation (see e.g. \cite{Padovani2015mnras}) for the all-flavor neutrino flux:
\begin{eqnarray}
E^2_\nu F_\nu(E_\nu) &=& \left[\int_{10\,\mathrm{GeV}}^\infty E_\gamma F_\gamma dE_\gamma\right] \nonumber\\&& {}\times
\frac{Y}{0.9} \left(\frac{E_\nu}{E_{\nu,\mathrm{peak}}}\right)^{1-s} 
\exp\left(-\frac{E_\nu}{E_{\nu,\mathrm{peak}}}\right),
\nonumber\\
\label{eq:blazarNu}
\end{eqnarray}
where $E_{\nu,\mathrm{peak}} \simeq 10$~PeV for typical 2FHL sources, and where $s = -0.35$ is adopted in order to obtain the denominator normalization factor of $0.9$. The $Y$ parameter absorbs the details of the actual particle interactions: The gamma-ray emission is mostly leptonic for $Y < 1$, and mainly due to synchrotron (from $p\pi$ interactions) when $Y \simeq 3$.

The model above features a very hard energy spectrum at the PeV scale.
We also test a phenomenological (hence less physically motivated) model where neutrino spectrum follows the gamma-ray spectrum, i.e., $E_\nu^{-2.5}$, as in the case of hadronuclear sources.
But in this case, in order not to violate the constraints from the diffuse gamma-ray background~\cite{ATZ,Murase:2013rfa}, we find that the blazar contribution at $\agt$10~TeV is extremely small.

\subsubsection{The isotropic component} 

Besides these physically-motivated models, we also consider an isotropic component with a power-law spectrum describing the flux from hypothetical additional sources currently not associated to a known point-source catalog in some wavelength.
We choose $P_\mathrm{iso}(I_\nu|E)$ to be normally distributed with a
fixed width $\mu/\sigma$, where $\mu$ and $\sigma$ are the mean and
rms of the intensity $I_\nu$, respectively. Although in principle a one-point fluctuation analysis would be sensitive to this information, we have checked that with current data our analysis is not sensitive to variations of the width $\mu/\sigma \in \{10,100,1000\}$. This model therefore has two adjustable parameters:
a normalization $\langle I_\nu\rangle_{100\;{\rm TeV}}$ and a power-law
spectral index $\Gamma$.
Since this component describes unknown sources, estimates for these parameters were
determined by the maximum likelihood method, and the estimate of their covariance
matrix by inversion of the observed Fisher
information~\cite{Edwards:2017mnf}.

We remark that these two are the {\it only} free parameters in our description of the high-energy neutrino sky; the parameters for all the other components have been determined using multi-messenger information.
}

\subsection{The Galactic components}\label{sec:CosmicRays}

A very relevant issue is the role of the Galactic contribution: Although there is currently no positive and statistically significant evidence for it, the expectation is that such a component should exist.
We consider here physically motivated models for the Galactic cosmic-ray contribution, some of which are further tuned to a high-energy extrapolation of gamma-ray observations: The {\it Canonical} and {\it Gamma} models,  {presented in Ref.~\cite{Gaggero2015ApJ} (see fig. 2 therein for a comparative plot)}.
These models are implemented with {\tt DRAGON}~\cite{Evoli2008JCAP}, a numerical package designed to simulate all processes related to cosmic-ray transport by solving a time-dependent diffusion-loss
equation for all the relevant species, and are all tuned to GeV--TeV local charged cosmic-ray data \cite{AMSproton2011,CREAMproton2011}.

\begin{figure}
\begin{center}
\includegraphics[width=8.5cm]{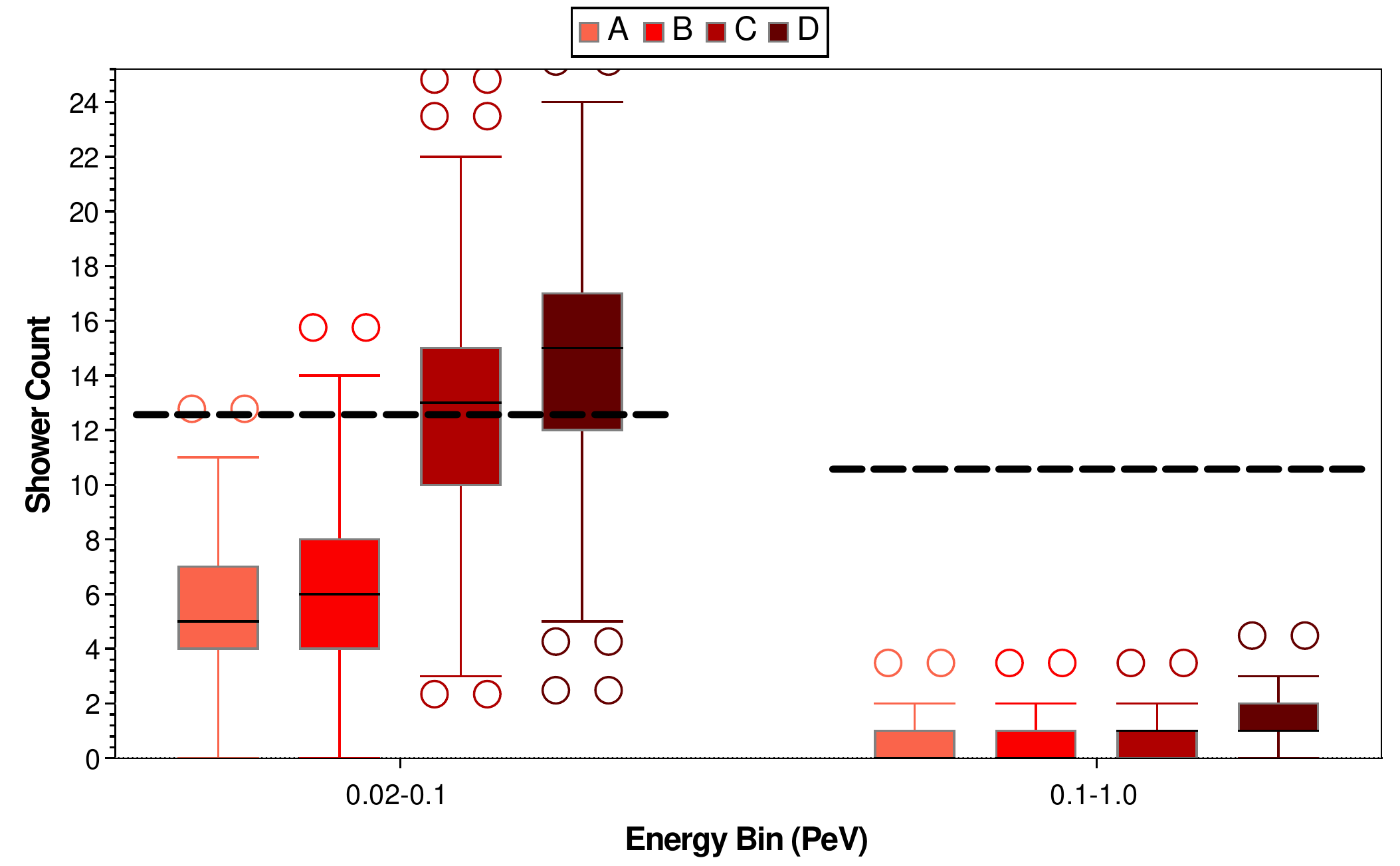}
\caption{Comparison of the count distributions of four model Galactic contributions (whiskered boxes) to the HESE events (dashed lines). Including other components would broaden the boxes and shift them to higher counts (cf. Fig.~\ref{fig:all_energies_boxplot}, left). Models \m{A} and \m{B} are an extrapolation of GeV--TeV local cosmic ray data, while models \m{C} and \m{D} are extrapolated from Fermi data (see main text, and Ref.~\cite{Gaggero2015ApJ}). The gamma-ray extrapolated models overpredict the IceCube data below 100~TeV. The signal region adopted in this figure contains only showers with energies below 1~PeV, and located either at declinations $-20^\circ < \delta < 20^\circ$ or within one pixel of the galactic center.}
\label{fig:galactic_boxplot}
\end{center}
\end{figure}

The {\it Gamma} models are also tuned to an extrapolation of Fermi-LAT gamma-ray data, as first discussed in Ref.~\cite{Gaggero2015PRD}. The key feature 
of those scenarios is a progressively harder proton spectrum in the inner Galaxy, {which shows a  progressive transition from a power law with index $\simeq -2.7$, inferred locally, to a harder one with index 
$\simeq -2.4$ at the GC: Such trend was recently confirmed by the Fermi-LAT collaboration, as shown e.g. in Fig. 8 of \cite{Acero:2016qlg}. This behavior is phenomenologically reproduced by means of a} transport scenario characterized by a harder scaling of the diffusion coefficient with rigidity in the inner Galactic plane (see Refs.~\cite{Cerri2017,Recchia2016MNRAS}  for two recent physical models that may result in this behavior).
As shown in Refs.~\cite{Gaggero2015ApJ,Gaggero2017PRL}, these models are characterized by a significantly larger gamma-ray flux in the $1$--$50$ TeV range, thus reproducing in a natural way the bright multi-TeV emission measured by the H.E.S.S. collaboration in the Galactic ridge region~\cite{HESS2016Nature}, and the anomalous spectral point provided by MILAGRO in a region of interest located in the inner Galactic plane~\cite{Milagro2008ApJ}. This also results in an increased neutrino flux at those energies, as seen in Fig.~\ref{fig:galactic_boxplot}.

Given the large uncertainty in the high-energy part of the spectrum, we consider two different realizations for both the {\it Canonical} and {\it Gamma} model,
characterized by different values for the proton high-energy injection spectrum cutoff.
The Galactic models are labeled as \{\m{$\emptyset$}, \m{A}, \m{B}, \m{C}, \m{D}\} for convenience.  Models \m{A} and \m{B} are {\it Canonical}, while models \m{C} and \m{D} are {\it Gamma}; models \m{A} and \m{C} are tuned to recent KASCADE data~\cite{Apel:2013uni} and feature a 5-PeV cutoff, while models \m{B} and \m{D} feature a very optimistic value of 50-PeV cutoff, again following Ref.~\cite{Gaggero2015ApJ}. We remark the very large uncertainty affecting the models in this high-energy range (see, e.g., the indications for a sub-PeV knee reported by the ARGO collaborations~\cite{Bartoli:2015vca}).
The model in which there is no Galactic contribution to the IceCube flux is labeled \m{$\emptyset$}.

For each of these models, a high-resolution map of the neutrino flux from Galactic cosmic rays, assuming a flavor ratio of 1:1:1, was produced using {\tt GammaSky}, a dedicated code developed by the {\tt
DRAGON} team~\cite{DiBernardo2013JCAP}. However, the large angular uncertainties associated with showers in IceCube data only gives us access to a low-resolution map. The probability distribution $P(F|\m{M})$ of this flux, in each pixel of the low-resolution map, was set to a Gaussian with parameters determined by the mean and variance of the oversampled map provided by {\tt GammaSky}. Additionally, these PDFs per pixel in different energy sub-bins were convolved into three energy bins~\cite{Feyereisen2016}, by treating the
spectrum produced by {\tt DRAGON} as a piecewise power law in each energy sub-bin and each pixel.

\section{Results}

We compute the likelihood for a model $\m{M}$ based on the total count distribution as
\begin{equation}
\mathcal{L} = \prod_{\mathrm{pixels}~p}~\prod_{\mathrm{bins}~E}  P(C=d^{(p,E)}|p,E,\m{M}),
\label{eq:likelihood}
\end{equation}
where the count distributions of different astrophysical and atmospheric
components of the model are combined by convolution, i.e., $P(C|\m{M}) = P(C|\m{M}_{\rm SFG}) \star P(C|\m{M}_{\rm BL-Lac}) \star P(C|\m{M}_{\rm prompt}) \star \cdots$. 
{The flux PDFs of starbursts and blazars are shown in Fig.~3 of Ref.~\cite{Feyereisen2016},  and the count PDF $P(C)$ is obtained by convoluting them with the Poisson distribution~\cite{Feyereisen2015}.
These PDFs are non-Poissonian with typically a power-law tail that makes the distribution skewed.
The intrinsic skewness, however, is dominated by shot noise of finite neutrino counts in contemporary neutrino telescopes. 
}

We add the best-fit isotropic contribution to this marginal likelihood as $\mathcal{L}_i = P(C|\m{M})\star P_\mathrm{iso}(C|i)$, where $i\in \{\m{\emptyset},\m{A} ,\m{B},\m{C},\m{D}\}$.

As summarised in Fig.~\ref{fig:all_energies_boxplot}, our main result is the strong evidence for a dominant isotropic component. The best-fit values of the normalization and spectral index for this component of unknown origin are presented in Table~\ref{tab:bestfit}, for all different choices of the Galactic template, including the null one.

\begin{table}[h]
\caption{Best-fit values, using six years of HESE shower data, for the
 isotropic components associated to each model of the Galactic
 contribution. The normalization at 100~TeV is quoted in units of
 $10^{-18}~\units$.}
\begin{tabular}{|c|ccc|}
\hline
Model & Normalization & Spectrum & (Correlation)\\
\hline
\m{$\emptyset$} & $3.42\pm0.22$ & $2.84\pm0.63$ & $-0.62$\\
\m{A} &  $2.86\pm0.22$ & $2.71\pm0.53$  & $+0.11$\\
\m{B} & $2.81\pm0.21$ & $2.71\pm0.54$ & $+0.18$ \\
\m{C} & $2.71\pm0.20$ & $2.69\pm0.56$ & $+0.32$ \\
\m{D} & $2.64\pm0.19$ & $2.69\pm0.58$ & $+0.41$ \\
\hline
\end{tabular}
\label{tab:bestfit}
\end{table}

All these models provide decent fits to the data. For instance, the (two-sided, pre-trials) $p$-values for models $(\m{A}, \m{C})$ are
$p\approx (0.54, 0.50)$.
The values for the best-fit spectra are all compatible with each other. However, there is a $3\sigma$ tension in normalization between models
with and without a Galactic contribution.
This shift of $\Delta\langle I_\nu\rangle\gtrsim 0.6 \times 10^{-18}
~\units$ is consistent with the typical normalization of the Galactic models at 100~TeV~\cite{Gaggero2015ApJ,Gaggero2017PRL}.

As far as the comparison between different Galactic scenarios is considered, the likelihood ratio $\Lambda$ can be used to compare these models amongst each other.
We quote all possible pairings $\Lambda$ in Table~\ref{tab:BayesFactors} in units of information (deci hartley; dH), i.e., as 

\begin{equation}
\left(\frac{\Lambda_{\m{a}\m{b}}}{1~\mathrm{dH}}\right) = 10\log_{10}\left(\frac{\mathcal{L}_\m{a}}{\mathcal{L}_\m{b}}\right).
\label{likelihoodRatio}
\end{equation}

In these units, $\Lambda_{\m{a}\m{b}} = 20$ would correspond to odds of $100:1$ in favour of model \m{a}. Table~\ref{tab:BayesFactors} shows that the high-energy extrapolations of the {\it Canonical} models \m{A} and \m{B} are favored over the {\it Gamma} models \m{C} and \m{D}. There is no significant preference between a cutoff at 5 or at 50 PeV. Given the large uncertainties associated with cosmic ray transport modeling in the TeV--PeV domain, these results need to be taken with a grain of salt, as discussed in the following section.

\begin{table}[h]
\caption{{Likelihood ratios (as defined in Eq.~\ref{likelihoodRatio})} obtained using six years of HESE shower events, and the best-fit isotropic component of each model. The high-energy extrapolations of {\it Canonical} models (\m{A}, \m{B}) are favored over {\it Gamma} models (\m{C}, \m{D}).}
\begin{tabular}{|cc|cc|cc|}
\hline
~$\Lambda_\m{AC}$~ & ~$\Lambda_\m{AD}$~ & ~$\Lambda_\m{BC}$~ &
 ~$\Lambda_\m{BD}$~&~$\Lambda_\m{AB}$~&~$\Lambda_\m{CD}$~\\
19.8 & 25.7 & 19.2 & 25.1 & 0.6 & 5.9 \\
\hline
\hline
~$\Lambda_\m{A\emptyset}$~ & ~$\Lambda_\m{B\emptyset}$~ &
	 ~$\Lambda_\m{C\emptyset}$~ & ~$\Lambda_\m{D\emptyset}$~&&\\
7.1 & 6.5 & $-12.7$ & $-18.6$ & &\\
\hline
\end{tabular}
\label{tab:BayesFactors}
\end{table}

\section{Discussion}

The first relevant discussion point is the nature of the isotropic component outlined by this analysis.
This template captures both the effect of mis-modeling  of the contributions we have included in the analysis, and the effect of astrophysical contributions yet to be considered. Assuming a {\it Canonical} model, the best-fit normalization of the isotropic
component is ${\langle I\rangle_{100\;\mathrm{TeV}}} = (2.8\pm0.2) \times 10^{-18}\;\units$ and its best-fit spectral index is $\Gamma=-2.7\pm0.5$.

The missing flux has a spectrum consistent with astrophysics; a missing contribution from atmospherics would have a spectrum closer to $E^{-3.7}$ (2$\sigma$ away from the best-fit value).
The normalization is four to five times larger than that of the Galactic contribution, consistent with the absence of positive evidence for such a subdominant contribution in this and other studies (e.g., \cite{Halzen:2016seh}).

There are not many source populations that are predicted to contribute to the IceCube flux significantly, yet have a spectrum softer than $E^{-2.7}$. The regular star-forming galaxies have a spectrum close to this, but
such a component is has already been studied in a one-point fluctuation analysis~\cite{Feyereisen2016}, and their
normalization is much too small to account for the entire isotropic component even allowing for systematic uncertainties.

It has also been suggested that radio galaxies could give significant contribution to most of the IceCube neutrino events~\cite{Hooper2016b}. The spectrum, however, appears to be much harder than $E^{-2.7}$ according to gamma-ray data~\cite{DiMauro:2013xta}. In any case, any transparent hadronuclear sources have to have spectra harder than $E^{-2.2}$ or so in order to give substantial contribution to the IceCube neutrinos, according to the Fermi diffuse gamma-ray background spectrum~\cite{Murase:2013rfa} and cross correlation
measurement~\cite{ATZ}.

These considerations naturally lead us to consider some hidden source class, where gamma rays cannot escape.
This includes both photohadron and hadronuclear processes in mildly relativistic or {choked jets~\cite{Tamborra:2015fzv,2016PhRvL.116g1101M, Senno:2015tsn}, for which some soft spectrum component can naturally arise without being constrained by the gamma-ray data: Such a scenario can be tested by looking for correlation with low-power gamma-ray bursts (exploiting future, more sensitive GRB satellites)}.

Let us now turn our attention to the results on the Galactic contribution.
The direction of the Pearson correlation between the best-fit normalizations and spectral index in Table~\ref{tab:bestfit} changes
between models with and without a Galactic contribution.
Since the normalization is taken at 100~TeV, the sign of the correlation tells us whether the isotropic component is mostly fitting data at lower energies
(negative) or at higher energies (positive).
We find that the isotropic component in model \m{$\emptyset$} is mostly trying to produce counts at low energies.
Meanwhile with the {\itshape Canonical} models the low-energy bins have more counts to start with, and so the isotropic component becomes more
relevant at higher energies.

There is a similar increase in the correlation coefficient between {\it Canonical} and {\it Gamma} models. Under this interpretation of the correlation coefficients, the likelihood ratios in Table~\ref{tab:BayesFactors} indicate that {\it Gamma} models are
over-predicting the neutrino flux in the lower energy bins of the analysis {(see also Fig. \ref{fig:galactic_boxplot})}.
This conclusion is also hinted at by the
lower normalizations associated to these models, as a consequence of the fact that
{\it Gamma} models produce a larger flux than {\it Canonical} models at these energies (cf., Fig.~\ref{fig:galactic_boxplot} and Refs.~\cite{Gaggero2015ApJ,Gaggero2017PRL}).

We remark that we are considering here energies larger than $20$~TeV, while the {\it Gamma} models are tuned to Fermi-LAT data only up to $\simeq 300$~GeV, and were shown to reproduce several other gamma-ray datasets in the $1$--$50$~TeV energy range as well.
If confirmed by the forthcoming data releases, the different indications coming from (mostly sub-TeV) gamma-ray data, which show strong preference for the {\it Gamma} models, and the TeV--PeV neutrino data that seem to favor the extrapolation of {\it Canonical} models, can reveal hints of either different transport regimes at work in different energy ranges, or different classes of Galactic sources at work.

\section{Summary}

We presented a one-point fluctuation analysis of the six-year high-energy starting events (HESE) shower data, based on the methodology and models presented in Ref.~\cite{Feyereisen2016}. Our comprehensive analysis included for the first time both a phenomenological isotropic component representing an additional extragalactic contribution with a Gaussian flux distribution and a power-law spectrum, and a set of predictions for the Galactic component obtained with the {\tt DRAGON} code. {We found that the additional isotropic template, not associated with well-measured point-source classes, actually dominates the neutrino sky, and we discussed possible ideas regarding its origin}. Our result is robust with respect to variations of the Galactic contribution. All the Galactic models considered here show decent fits of the data, with preference for the high-energy extrapolation of {\it Canonical} models with respect to the more optimistic {\it Gamma} models, tuned on the large-scale trends inferred by Fermi-LAT diffuse data. This result shows the power of this kind of analysis in placing constraints on phenomenological Galactic cosmic ray transport models. If confirmed, the presence of conflicting indications coming from gamma-ray data in the GeV--TeV domain with respect to those coming from neutrino data in the TeV--PeV range can provide indications on the physics of cosmic ray transport in the high-energy regime.

\acknowledgments
We thank D. Grasso and A. Marinelli for inspiring discussions.
This work was supported by the Netherlands Organization for Scientific Research (NWO) through Vidi grant (MRF and SA), and also in part by JSPS KAKENHI Grant Number JP17H04836 (SA).

\appendix
\section{Predictive one-point fluctuation analysis}

\begin{figure}
\begin{center}
\includegraphics[width=8.5cm]{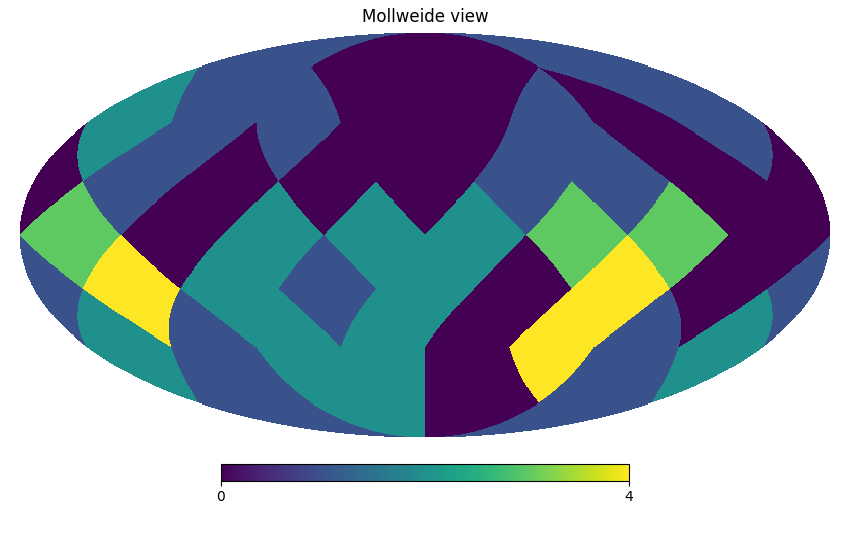}
\caption{HealPix visualisation of six years of HESE Showers (in decl/RA coordinates) used in this analysis.}
\label{fig:Healpix_map}
\end{center}
\end{figure}

{Ref.~\cite{Feyereisen2016}, which we expand upon in this study, describes a methodology for turning multimessenger data into a predictive model of the neutrino count distribution in IceCube, and reports on the first one-point analysis of the HESE data. In this appendix, we briefly review this methodology.

The neutrino intensity due to any class of astrophysical sources incident on IceCube, can be computed using extrapolations of multimessenger data. Particularly, given some model $M$ of these sources' neutrino flux and redshift distributions, the probability distribution of their intensity $P(I_\nu|M)$ can be computed with no additional assumptions \cite{Feyereisen2015}. The total intensity $I = I_1+\cdots+I_n$ due to multiple classes of astrophysical source with models $M_1, \cdots, M_n$ is then distributed as \begin{equation}
P(I|M_\mathrm{tot}) = P(I|M_1) \star \cdots \star P(I|M_n)
\end{equation} where $\star$ denotes convolution. Backgrounds to our signal (such as atmospheric backgrounds), individual and possibly extended sources (such as the Milky Way), and phenomenological contributions (such as the one considered in the main text), provided they can be described by an intensity distribution, may similarly be convolved into the distribution for the total intensity. It is understood that this neutrino intensity distribution is a function of the line of sight and energy of this incident intensity, incorporating both spectral and anisotropic features.

The count distribution is then constructed as the discretisation of the flux distribution, accounting for the exposure $\epsilon$ of the detector:\begin{equation}
P(C|M_\mathrm{tot}) = \int \mathcal{P}(C|I \times \epsilon) P(I|M_\mathrm{tot}) dI.
\label{eq:count_integral}
\end{equation} The IceCube exposure $\epsilon$ also has spectral and anisotropic features, which depend both on the geometry and location of the detector, and on the vetos imposed on the HESE analysis. These will be discussed in the second part of this Appendix.

A full model of the detector would allow us to account for the measurement uncertainties on the energy and the line of sight of individual events when translating this predictive count distribution into the likelihood function of the data. If we denote collectively by $\psi$ the observing energy and line of sight, and by $P(\sigma|\psi)$ the energy resolution and angular resolution (as a function of $\psi$), then the likelihood of any HESE event $d(\psi)$ is a weighted convolution of the $P(C)$ of all $\psi$ within the instrumental resolution. We have: \begin{equation}
\ell(d) = \left. \convint P(C|M,\psi+\sigma) P(\sigma|\psi) d\sigma \right|_{C=d}
\end{equation} where ${\mathop{\mathrlap{\;\star}\int}}$ was defined in Ref.~\cite{Feyereisen2016}. The (unbinned) likelihood of the HESE data is then a product of $\ell(d)$'s.

However, to simplify the analysis, we instead bin the HESE shower data into three large energy bins, and into pixels large enough to fully contain the the angular resolution.\footnote{This binning scheme effectively sets $P(\sigma|\psi)$ to a multi-dimensional rectangle function, rather than the multi-dimensional Gaussian one might naively expect. This is not too much of a problem for isotropic contributions, but it is not clear how much this affects the Galactic contribution in our model.} The count distribution $P(C|M_\mathrm{tot})$, in each pixel and each energy bin, therefore constructs directly the binned likelihood $\mathcal{L}$ presented in the main text. The pixel binning scheme (Healpix) is presented in Figure \ref{fig:Healpix_map}.}

\section{Characterisation of IceCube}

{In this appendix, we briefly review the treatment of the IceCube detector adopted in Ref.~\cite{Feyereisen2016}, which we adopt in this study.

The effective area relevant to the HESE shower analysis is declinantion, energy, and flavour dependent. The IceCube Collaboration provides a separate estimation of the effective area for each of the three flavours, which we interpolate in declination and energy. The effective area for showers depends on the probability $p_{S}^{e/\mu/\tau}$ that a neutrino of a given flavour (sampled randomly from the total neutrino flux) produces a shower. Assuming an equal abundance of neutrinos and antineutrinos, we use the approximation
$p^\mu_S = 0.2,~p^{e/\tau}_S = 1$ \cite{Palladino} to write
\begin{equation}
A_\mathrm{eff} = 2 \sum_{f\in\{e,\mu,\tau\}} p^f_{S} \times A^f \times \eta^f\ ,
\end{equation} 
where $A^f$ is the flavour-energy-and-declination dependent quantity given by IceCube and $\eta^f$ is the fraction of neutrinos of a given flavour ($\eta= 1/3$ for a $1:1:1$ flavour ratio). We employ a $1:1:1$ ratio for all extragalactic components, and a $1:1:0$ ratio for the prompt atmospheric flux \cite{Enberg:2008te}. We approximate the average conventional atmospheric flux given by Ref.~\cite{Honda:2015fha} as $1.77\times 10^{-14}\;\units$ with a flavour ratio of $1:35:0$.\footnote{We have explicitly checked that the results are not significantly affected by using the model from Ref.~\cite{Feyereisen2016}, which assumed fewer significant digits.} Other percent-level atmospheric contributions from $\nu_e$ and $\nu_\tau$ fluxes \cite{Honda:2015fha,Enberg:2008te} are neglected, as are the neutrino-antineutrino ratios, although the fully detailed (even energy-dependent) flavour ratios can in principle be accounted for in this type of analysis.

Computing the predicted number of events $I\times \epsilon$ in Eqn.~\eqref{eq:count_integral} relies on this effective area. Since we are interested in the number of counts per pixel and per energy bin, we are effectively computing the integrated event counts; however, to compute the distribution of the integrated counts, we must compute the probability distribution of an integral. The ${\mathop{\mathrlap{\;\star}\int}}$ operation defined in Ref.~\cite{Feyereisen2016} can, by representing this integral as a Riemann summation, be approximated by a large but finite number of convolutions -- the interpolation of $A_\mathrm{eff}(E)$ is then an essential ingredient to compute the probability distribution of the integrated event counts.

This study departs from Ref.~\cite{Feyereisen2016} only minimally in its treatment of the declination dependence of the effective area. As discussed above, we simplify the analysis by binning the HESE data into pixels with HealPix \cite{HealPix}. Although the effective area does vary across these large pixels, it varies monotonically, so we use the value at the central declination of each pixel to approximate $A_\mathrm{eff}$. Unlike Ref.~\cite{Feyereisen2016}, we do not exploit the isolatitudinality of HealPix pixels and the effective area to accelerate our analysis, in order to accomodate for the symmetry-breaking contribution from the Galactic plane.}

\bibliographystyle{utphys}
\bibliography{refs}

\end{document}